\documentclass[10pt, conference, compsocconf]{IEEEtran}

\usepackage{cite}
\usepackage{graphicx}
\usepackage{algorithmic}
\usepackage{color}
\usepackage{xcolor}
\usepackage{url}
\usepackage{comment}
\usepackage{amssymb}
\usepackage{pifont}
\usepackage{colortbl}
\usepackage{booktabs}
\usepackage{setspace}
\usepackage{footnote}
\usepackage{amsbsy}

\definecolor{light-gray}{gray}{0.93}

\begin{document}
\title{DroidAnalytics: A Signature Based Analytic System to Collect,
Extract, Analyze and Associate Android Malware}

\author{Min Zheng, Mingshen Sun, John C.S. Lui\\
Computer Science \& Engineering Department\\
The Chinese University of Hong Kong
}
\maketitle

\begin{abstract}
Smartphones and mobile devices are rapidly becoming
indispensable devices for many users.
Unfortunately, they also become
fertile grounds for hackers to deploy malware and to spread virus.
There is an urgent need to have a ``{\em security analytic \& forensic system}''
which can facilitate analysts to examine, dissect, associate and correlate
large number of mobile applications.
An effective analytic system needs to
address the following questions:
{\em
How to automatically collect and manage a high volume of mobile malware?
How to analyze a zero-day suspicious application, and compare or
associate it with existing malware families in the database?
How to perform information retrieval so to reveal
similar malicious logic with existing malware, and to quickly identify
the new malicious code segment?
}
In this paper, we
present the design and implementation of
{\em DroidAnalytics}, a signature based analytic
system to automatically collect, manage, analyze and extract android malware.
The system facilitates analysts to retrieve,
associate and reveal malicious logics at the ``{\em opcode level}''.
We demonstrate the efficacy of DroidAnalytics using 150,368
Android applications,
and successfully determine
2,494 Android malware from 102 different families, with
342 of them being {\em zero-day} malware samples from six different families.
To the best of our knowledge, this is the first reported case
in showing such a large Android malware analysis/detection.
The evaluation shows the
DroidAnalytics is a valuable tool and is effective
in analyzing malware repackaging and mutations.
\end{abstract}

\section{{\bf Introduction}} \label{section: introduction}

Smartphones are becoming prevailing devices for many people.
Unfortunately, malware on
smartphones is also increasing at an unprecedented rate.
Android OS-based systems, being the most popular platform
for mobile devices, have been a popular
target for malware developers.
As stated in \cite{mcafee:mcafee}, the exponential growth of mobile
malware is mainly due to the ease of generating malware variants.
Although there are number of works which focus on Android malware detection
via permission leakage,
it is equally important to
design a system that can perform comprehensive {\em malware analytics}:
analyze and dissect suspicious applications at the {\em opcode level}
(instead at the permission level),
and correlate these applications to existing malware in the database
to determine whether they are mutated malware or even zero-day malware,
and to discover which legitimate applications are infected.

\noindent {\bf Challenges:}
To realize an effective analytic system
for Android mobile applications,
we need to overcome several technical hurdles.
First, how to systematically {\em collect} malware from the wild.
As indicated in \cite{blogspot},
new malware variants are always hidden
in many different third-party markets.
Due to the competition of anti-virus companies and their fear
of accidentally releasing malware to the public,
companies are usually reluctant to share their malware database to researchers.
Researchers in academic can only obtain a small number of
mobile malware samples.
Hence, how to {\em automate a systematic process} to
obtain these malicious applications
is the first hurdle we need to overcome.

The second hurdle is
how to identify {\em repackaged applications} (or mutated malware)
from the vast ocean of applications and malware.
As reported in\cite{dimva12},
hackers can easily transform legitimate applications
by injecting malicious logic or obfuscated program segments
so that they have the same structure as the original application but
contain malicious logic.
Thus, how to determine whether an application is a repackaged
or obfuscated malware, and which legitimate applications are infected
is very challenging.

The third hurdle is how to {\em associate} malware
with existing malware (or application) so as to facilitate security analysis.
The existing approach of using cryptographic hash or package name as an
identifier is not effective because hackers can easily change
the hash value or package name.
Currently, security analysts need to
go through a laborous process of manually reverse engineer a malware
to discover malicious functions and structure.
There is an urgent need to have an efficient method
to associate malware with other malware in the database,
so to examine their commonalities at the opcode level.

\noindent
{\bf Contributions:}
To address these problems mentioned above,
we present the design and implementation of
{\em DroidAnalytics}, an Andorid  malware analytic system for
malware collection, signature generation,
information retrieval, and malware association based
on similarity score.  Furthermore,
DroidAnalytic can efficiently detect
zero-day repackaged malware.
The contributions of our system are:
\begin{itemize}
\item DroidAnalytics automates the processes of
malware collection, analysis and management. We have successfully collected
150,368 Android applications, and determined
2,494 malware samples from 102 families. Among those,
there are 342 zero-day malware samples
from six different malware families.
We also plan to release the malware database to
the research community (please refer to
\url{https://dl.dropbox.com/u/37123887/malware.pdf}).

\item DroidAnalytics uses a {\em multi-level signature algorithm} to extract the malware
feature based on their semantic meaning at the
{\em opcode level}.
This is far more robust than a cryptographic hash of the entire application.
We show how to use DroidAnalytics to combat
against malware which uses repackaging or code obfuscation,
as well as how to analyze malware with dynamic payloads
(see Sec.~\ref{sec: signature}).

\item Unlike previous works which associate
malware via ``{\em permission}'', DroidAnalytics
associates malware and generates signatures at the app/class/method level.
Hence, we can easily track and analyze mutation, derivatives,
and generation of new malware.
DroidAnalytics can
reveal malicious behavior at the method level
so to identify repackaged malware,
and perform class association among malware/applications
(see Sec.~\ref{sec: analytic_capability}).

\item We show how to use DroidAnalytics to detect
{\em zero-day} repackaged malware.
We have found
342 zero-day repackaged malware in six different families
(see Sec.~\ref{sec: zeroday}).

\end{itemize}

\section{{\bf Design \& Implementation of DroidAnalytics}} \label{design}

Here, we present the design and implementation of DroidAnalytics.
Our system consists of modules for
automatic malware collection,
signature generation,
information retrieval and association, as well as
similarity comparison between malware.
We will also show how to use these functions to
detect zero-day repackaged malware.

\begin{figure}[htb]
\begin{center}
\includegraphics[width=210pt]{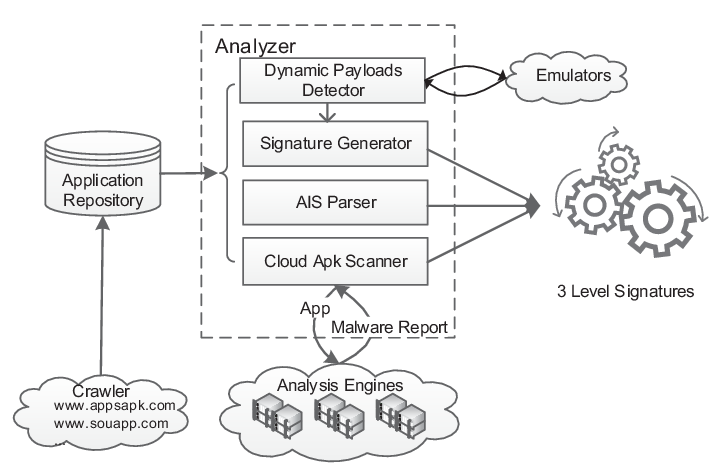}
\end{center}
\caption{The Architecture of the DroidAnalytics}
\label{figure: architecture}
\end{figure}

\subsection{Building Blocks of DroidAnalytics}

Figure~\ref{figure: architecture} depicts the architecture of DroidAnalytics
and its components.
Let us explain the design of each component.

\noindent
{\bf $\bullet$ Extensible Crawler:}
In DroidAnalytics, we implement an application crawler based
on Scrapy\cite{scrapy}. Users can specify
official or third party market places,
as well as blog sites and the crawler will perform
regular mobile application download.
The crawler enables us to systematically build up the
mobile applications database for malware analysis
and association.
So far, we have collected 150,368 mobile applications and carried out
detailed security analysis.

\noindent
{\bf $\bullet$ Dynamic Payloads Detector:}
To deal with malware which
dynamically downloads malicious codes via the Internet or attachment files,
we have implemented the dynamic payloads detector component, which
determines malicious trigger code within
malware packages and tracks the downloaded application and its behavior
in virtual machine.
Firstly, it scans the package to find suspicious
files such as {\tt .elf} or {\tt .jar} file.
Hackers usually camouflage
malicious files by changing their file type.
To overcome this, this component scans all files and identifies files
using their {\em magic numbers} instead of file extension.
Secondly, if an application has any
Internet behavior (e.g., Internet permission
or re-delegating other applications to download files\cite{redelegate}),
the dynamic payloads detector will treat these files as the target,
then runs the application in the emulator.
The system will use the forward symbolic execution technique
to trigger the download behavior.
Both the suspicious files within the package and
dynamically downloaded files from the Internet will be sent to
the signature generator (which we will shortly describe) for further analysis.

\noindent
{\bf $\bullet$ Android App Information (AIS) Parser:}
AIS is a data structure within DroidAnalytics  and it is used
to represent {\tt .apk} information structure.
Using the AIS parser, analysts can reveal
the cryptographic hash (or other basic signature) of
an {\tt.apk} file, its package name, permission information, broadcast
receiver information and disassembled code, $\ldots$, etc.
Our AIS parser decrypts the {\tt AndroidManifest.xml} within an application and
disassembles the {\tt .dex} file into {\tt.smali} code.
Then it extracts package information from source code and retains
it in AIS so analysts can easily retrieve this information.

\noindent
{\bf $\bullet$ Signature Generator:}
Anti-virus companies usually use cryptographic hash, e.g., MD5,
to generate a signature for an application.  This has two major drawbacks.
Firstly, hackers can easily mutate an application and change
its cryptographic hash.
Secondly, the cryptographic hash
does not provide sufficient flexibility for security analysis.
In DroidAnalytics, we use a {\em three-level} signature generation
scheme to identify each application.
This signature scheme is based on the mobile application, classes,
methods, as well as  malware's dynamic payloads (if any).
Our signature generation is based
on the following observation:  {\em For any functional application,
it needs to invoke various Android API calls,
and Android API calls sequence within a method is difficult to modify}
(unless one drastically changes the program's logic, but we did not find any
from the 150,368 applications we collected that used
this obfuscation technique).
Hence, we generate a method's signature using the API call sequence,
and given the signature of a method, create the signature of a class
which composes of different methods. Finally,
the signature of an application is composed of all signatures of its classes.
We like to emphasize that our signature algorithm
is not only for defense against malware obfuscation,
but more importantly, facilitating malware analysis
via class/method association (we will show in later sections).
Let us present the detail of signature generation.

\noindent
{\bf (a) Android API calls table:}
Our system uses the API calls table of the Android SDK.
The {\tt android.jar} file is the framework package provided by the Android SDK.
We use the Java reflection\cite{mccl98} to obtain all descriptions of
the API calls. For each API,
we extract both the {\em class path} and the {\em method name}.
We assign each full path method a hex number as part of the ID.
For the current version of DroidAnalytics, we extract 47,126 full path methods
in the Android SDK 4.1 version as our API calls table.
Table~\ref{table:APItable} depicts a {\em snapshot}
of API calls table, e.g.,
{\tt android/content/Intent;-><init> } is assigned an ID 0x30291.

\begin{table}[htb]
\begin{center}
{\scriptsize
\begin{tabular}{|l|l|}
\hline
{\bf Full Path Method} & {\bf Method ID} \\ \hline  \hline
{\tt android/accounts/Account;-><init> }&  0x00001 \\  
\mbox{\hspace{1.2in} $\vdots$}  & \mbox{\hspace{0.2in}$\vdots$} \\ 
{\tt android/content/Intent;-><init> }&  0x30291 \\  
{\tt android/content/Intent;->toUri } &  0x30292 \\  
{\tt android/telephony/SmsManager;->getDefault } &  0x39D53 \\ 
{\tt android/app/PendingIntent;->getBroadcast } &  0xF3E91 \\
\hline
\end{tabular}
}
\caption{Example of the Android API Calls Table and assigned IDs}
\label{table:APItable}
\end{center}
\end{table}

\noindent
{\bf (b) Disassembling process:}
Each Android application is composed of different classes
and each class is composed of different methods.
To generate signatures for each class or method, DroidAnalytics first
disassembles an {\tt .apk} file, then takes
the Dalvik opcodes of the
{\tt.dex} file  and transforms them
to methods and classes.
Then DroidAnalytics uses the Android API calls table to generate signatures.

\begin{figure}[htb]
\centering
\includegraphics[width=250pt]{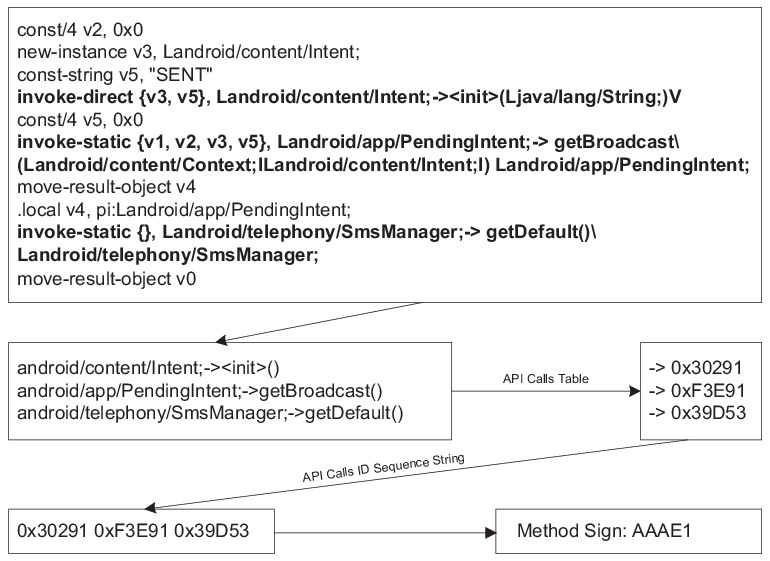}

\caption{The Process of Lev3 Signature Generation}
\label{Fig:lev3}

\end{figure}

\noindent
{\bf (c) Generate Lev3 signature (or method signature):}
The system first generates a signature for each method and we call this the
{\em Lev3 signature}.
Based on the Android API calls table,
the system extracts the API call ID sequence as a string in each method,
then hashes this string value
to produce the method's signature.
Figure~\ref{Fig:lev3} illustrates how to generate the
Lev3 signature of a method which sends messages to another mobile phone.
Figure~\ref{Fig:lev3} shows that the method contains
three API calls. Using the Android API calls table (as in
Table~\ref{table:APItable}), we determine their IDs.
Signature of a method is generated by cancatenation of all these IDs.
Note that
DroidAnalytics will not extract the API calls which will not be executed
in run time because these codes are usually generated via obfuscation.
Furthermore, if a method (except the main method) will not be
invoked by any other methods,
signature generator will also ignore this method because this may be a
defunct method generated by malware writers.

\noindent
{\bf(d) Generate Lev2 signature
(either  class signature or dynamic payload signature):}
Next, DroidAnalytics proceeds to generate the Lev2 signature for each {\em class}, and it is
based on the Lev3 signatures of methods within that class.
Malware writers may use various obfuscation or repackaging
techniques to change the {\em calling order} of the methods table in
a {\tt .dex} file. To overcome this
problem, our signature generation algorithm will first
{\em sort} the level 3 signatures within that class,
and then concatenate all these level 3 signatures to form
the level 2 signature.

Some malicious codes are dynamically downloaded from
the Internet during execution.
DroidAnalytics uses
the {\em dynamic payloads detector} component to obtain the payload files.
For the dynamic payloads which are {\tt .dex} file
or {\tt .jar} file, DroidAnalytics
treats them as classes within the malware.
Given these files, the system checks their API call sequence
and generates a
Lev2 signature for each class within an application.
For the dynamic payloads which contain, say, {\tt .elf} file or {\tt .so} file,
DroidAnalytics treats them as a single class
within that malware, then uses the cryptographic hash
value (e.g., MD5) of the payload as its Lev2 signature.
For the dynamic payloads which are {\tt .apk} files, DroidAnalytics
treats each as a new application and a class within the malware.
DroidAnalytics first
uses the cryptographic hash value (e.g., MD5) of the new {\tt .apk} file
as one Lev2 signature of that malware.
Because the payload is a new application,
DroidAnalytics will use the method we discussed
to carry out a new signature generation.

\noindent
{\bf (e) Generate Lev1 signature (or application signature):}
The Lev1 signature is based on the level 2 signatures,
e.g., signatures of all qualified classes within an application.
In addition, the signature generator will ignore the class
(except the main class)
which will not be invoked by any other classes
since these defunct classes may be generated via obfuscation.
Malware writers may use some repackaging or obfuscation techniques to
change the order of the classes table of the {\tt.dex} file, our
signature algorithm will first {\em sort} all Lev2 signatures,
then concatenate these Lev2 signatures to generate the Lev1 signature.

\begin{figure}[htb]
\centering
\includegraphics[width=210pt]{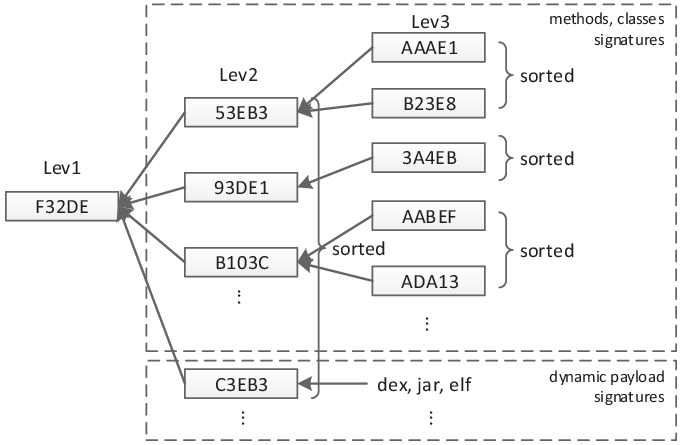}

\caption{Illustration of signature generation:
   the application (Lev1) signature, class level (Lev2) signatures
   and method level (Lev3) signatures.}
\label{Fig:SigAlg}
\end{figure}

Figure~\ref{Fig:SigAlg} summarizes the framework of our signature algorithm.
For example, the Lev3 signatures of {\tt AAAE1} and {\tt B23E8}
are the two method signatures within the same class. Based on these two
(sorted) signatures, we generate the Lev2 signature of the corresponding class,
which is {\tt 53EB3}.
Note that the Lev2 signature of {\tt C3EB3} is generated
from a {\tt .dex} file which is a dynamic payload used to execute
the malicious behavior.
Based on all sorted Lev2 signatures of all classes,
we generate the Lev1 signature, {\tt F32DE}, of the application.

For the current DroidAnalytics platform, we use a
server which is of 2.80 GHz Duo CPU processor,
4GB memory and 2 TB hard disk,
with two virtual machines in the server to implement the
anti-virus engine.
We carried out experiment to study the processing time to
scanning and generating signatures.
On average, it takes around 60 seconds
to scan one application (includes the dynamic analysis),
and around three seconds to generate all
three level signatures, five seconds to generate AIS information,
and one second to insert information into the database.
As of November 2012, our system have
downloaded 150,368 mobile applications from
the following places:
Google Play\cite{google:play},
nine Android third party markets (e.g., \cite{russia, appchina, souapp}),
two malware forums \cite{groups1, groups2}
and one mobile malware share blog\cite{blogspot}.
The size of all downloaded application is 468GB.

\section{{\bf Utility \& Effectiveness of
Signature Based System}} \label{sec: signature}

Here, we illustrate how DroidAnalytics' signatures can be used
to analyze (and detect) malware repackaging,
code obfuscation and malware with dynamic payloads.

\noindent
{\bf A. Analyzing Malware Repackaging}
Repackaging obfuscation technique is commonly used by malware writers
to change the cryptographic hash value of an {\tt .apk} file without modifying
the opcodes of the {\tt .dex} file.
This technique is different from the repackaged technique
which is to inject new packages into the legitimate applications.
For example, using {\sf Jarsigner} utility of Java SDK to
re-sign an {\tt .apk} file only changes the signature
part of one {\tt .apk} file,
and generates a new {\tt .apk} file which preserves the same logic and
functionality as the original one.
Another example is using the {\sf Apktool}\cite{apkt12},
which is a reverse engineering tool to disassemble and rebuild an
{\tt.apk} file without changing any assembly code.
Although there is no modification on
the assembly codes,
the recompiler may change the classes order and
methods order during the {\em recompiling process}.
Therefore, repackaging obfuscation is often used to
{\em mutate} an existing malware to generate a new version with
a different signature.
If an anti-virus system only identifies malware
based on a cryptographic hash signature,
then repackaging obfuscation techniques can easily
evade the detection.

DroidAnalytics can detect malware which is generated by
repackaging obfuscation.
Since there is no modification on the opcodes within the {\tt .dex} file,
and DroidAnalytics first sorts Lev2 and Lev3 signatures before generating
the Lev1 signature.
Therefore, DroidAnalytics will generate the {\em same} signature
as the original even when one repackages the {\tt .apk} file.

\begin{figure*}[htb]
\begin{center}
\fcolorbox{gray}{white}{
\includegraphics[width=0.8\textwidth]{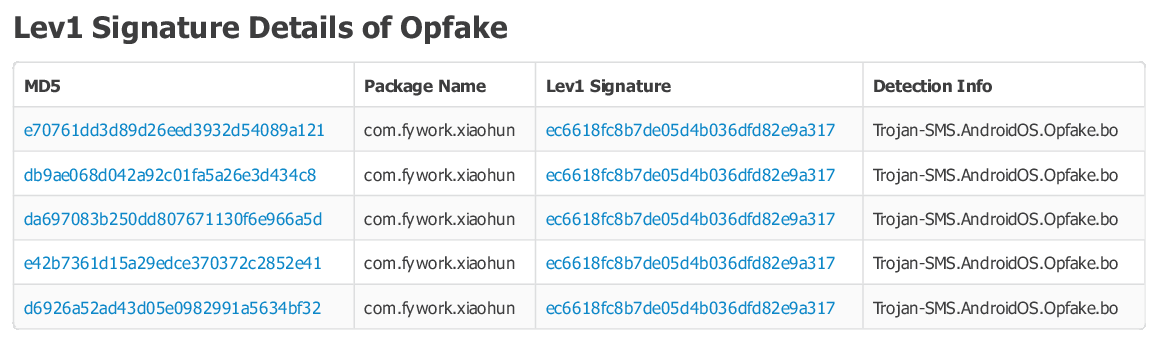}
}
\end{center}
\caption{Screen Capture of Opfake Family Lev1 Signature}
\label{Fig:lev1_opfake}
\end{figure*}

\noindent {\bf Experiment.}
To illustrate the above claim, we carry out the following experiment
and Figure \ref{Fig:lev1_opfake} illustrates the results.
{\sf Opfake} is a server-side polymorphism malware.
The malware mutates automatically when it is downloaded.
When analysts compare the cyclic redundancy codes (CRCs) of two
{\sf Opfake} downloads, it shows that the only meaningful
change happens in the file {\tt data.db}
which is located in ``{\tt res/raw/}'' folder.
The modified {\tt data.db} changes the signature data for the package
in ``{\tt META-INF}'' folder.
By analyzing this form of malware,
we find that all mutations in {\sf Opfake} family
happen in the same opcode (stored in {\tt classes.dex}).
Hence, our signature system will generate the
{\em same} level 1 signature for all mutations in this malware family.

\begin{figure*}[htb]
\begin{center}
\fcolorbox{gray}{white}{
\includegraphics[width=0.8\textwidth]{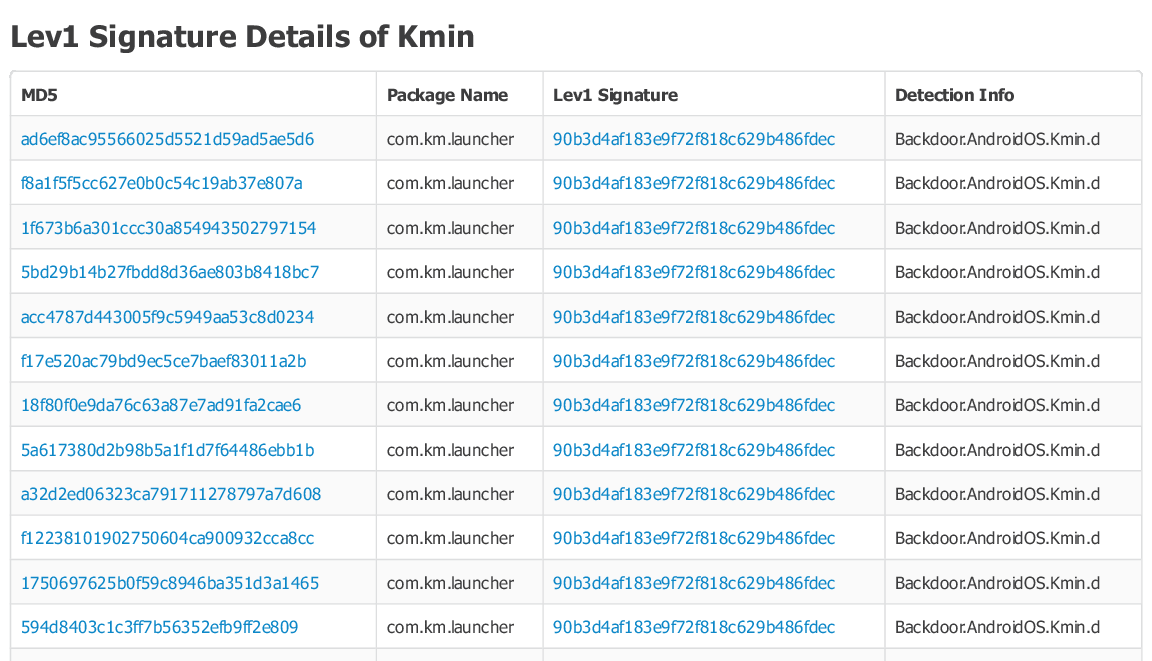}
}
\end{center}
\caption{Screen Capture of Kmin Family Lev1 Signature}
\label{Fig:lev1_kmin}
\end{figure*}

Figure \ref{Fig:lev1_kmin} illustrates another example
of using DroidAnalytics to analyze the {\sf Kmin} family.
We first calculate the
Lev1 signatures of all
150,368 applications in our database. After the calculation,
the result shows that the most frequent signature
(which corresponds to the
Lev1 signature {\tt 90b3d4af183e9f72f818c629b486fdec})
comes from 117 files and all these files have different MD5 values.
This shows that conventional cryptographic hashing (i.e., MD5) cannot identify
malware variants but DroidAnalytics can effectively identify them.
Also, these 117 files are all variants of the {\sf Kmin} family.  After
further analysis, we discover that the {\sf Kmin} family is a
wallpaper changer application, and all its variants have
the same application structure and same malicious behavior.
The only difference is that they have different icons and wallpaper files.

\noindent
{\bf B. Analyzing Malware which uses Code Obfuscation:}
A malware writer can use a
disassembler (e.g., {\sf Apktool}\cite{apkt12})
to convert a {\tt .dex} file into {\tt .smali} files,
then injects new malware logic into
the {\tt .smali} code, rebuilds it back to a {\tt .dex} file.
Based on this rebuilt process,
malware writers can apply various code obfuscation techniques
while preserve the
behavior as the original one in order to bypass the anti-virus detection.
As shown in~\cite{dimva12,Andreas07}, many mobile
anti-virus products are not effective to detect code obfuscated variants.

DroidAnalytics
will not extract the API calls in methods and classes
which will not be executed in run time (refer to Section \ref{design})
because they are defunct and can be
generated by obfuscators. In addition, the signature generation
does not depend on the name of methods or classes, hence
name obfuscation has no effect on our signature.
Furthermore, the signature generation of
DroidAnalytics is based on the analyst-defined API calls table.
So one can flexibly update the table entries to
defend against various code obfuscation techniques.

\noindent {\bf Experiment.}
To illustrate the effectiveness of DroidAnalytics against code obfuscation,
we chose 30 different malware from
three different families (10 samples in each family)
and Table \ref{table:obfs} illustrates our results.
The malware families in our study are:
{\sf Basebrid} (or {\sf Basebridge}), {\sf Gold-Dream},
and {\sf Kungfu}. We use ADAM\cite{adam12}, a system which
can automatically transform an original malware sample to different
variants by various repackaging and obfuscation
techniques (e.g. inserting defunct methods,
modifying methods name, ...etc). We
generate seven different variants for each malware. Then we put these
240 new malware samples into the DroidAnalytics system and check
their signatures. After our signature calculations,
the result shows that for each malware, the original sample and seven
mutated variants have {\em distinct} MD5 hash values
(3 repackaging, 4 code obfuscation),
but all of them have the {\em same level 1 signature}.
This shows that DroidAnalytics' signature system
is effective in defending against code obfuscation.

\begin{table*}[htb]
\centering
{\footnotesize
\begin{tabular}{|l|l|l|l|l|} \hline \hline
{\bf MD5} & {\bf Pakage Name} & {\bf Lev1 Signature} & {\bf Malware Family} & {\bf Description}\\ \hline
f7967f71b2f32287660ae9a3fa366022 & com.tutusw.phonespeedup & 3c83ed0f80646c8ba112cb8535c293dd & Kungfu & Original malware\\
f2e2d727f95fa868fd7ff54459e766e3 & com.tutusw.phonespeedup & 3c83ed0f80646c8ba112cb8535c293dd & Kungfu & Repackaging\\
e01f573cca83fdf2737be6ecee35fe33 & com.tutusw.phonespeedup & 3c83ed0f80646c8ba112cb8535c293dd & Kungfu& Repackaging\\
d383ceeb9c6ffcff8c0dd12b73ec43e3 & com.tutusw.phonespeedup & 3c83ed0f80646c8ba112cb8535c293dd & Kungfu& Repackaging\\
cd040541693693faca9fec646e12e7e6 & com.tutusw.phonespeedup & 3c83ed0f80646c8ba112cb8535c293dd & Kungfu& Obfuscation\\
401952d745cd7ca5281a7f08d3e2eede & com.tutusw.phonespeedup & 3c83ed0f80646c8ba112cb8535c293dd & Kungfu& Obfuscation\\
271c3965c7822ebf944feb8bbd1cfe7f & com.tutusw.phonespeedup & 3c83ed0f80646c8ba112cb8535c293dd & Kungfu& Obfuscation\\
8b12ccdc8a69cf2d6a7e6c00f698aaaa & com.tutusw.phonespeedup & 3c83ed0f80646c8ba112cb8535c293dd & Kungfu& Obfuscation\\
\hline \hline
\end{tabular}
\caption{Examples of Code Obfuscation}
\label{table:obfs}
}
\end{table*}

\noindent
{\bf C. Analyzing Malware with Attachement Files or Dynamic Payloads:}
Some malware will dynamically download file
which contains the malicious code from the Internet.
Also, some attachment files within a package
may contain malicious logic but they can be concealed
as other valid documents (e.g., {\tt .png} file, {\tt .wma} file).
DroidAnalytics will treat these files as dynamic payloads.
By using both static and dynamic analysis techniques
described in Section \ref{design}, DroidAnalytics accesses these payloads
and generates different signatures.

\noindent {\bf Experiment.}
We carried out the following experiment.  From our
malware database, we used our signature
system and detected some malware
contain the same file with a {\tt .png} filename extension.
But when we check the magic number of this file, it is actually
an {\tt .elf} file. Upon further analysis,
we found that this file is a root exploit and this malware belongs to
the {\sf GinMaster} (or {\sf GingerMaster}) family.
Another example is the {\sf Plankton} family. By using dynamic analysis,
DroidAnalytics discovered that all malware in this family will
download a {\tt plankton\_v0.0.4.jar} (or similar {\tt .jar})
when the main activity of the application starts.
Further analysis revealed the {\tt .jar} file
contains malicious behavior, i.e., stealing browser's history
information, making screen shortcuts and botnet logic.
Table \ref{table:dynamic_payload} depicts DroidAnalytics
system detects some representative malware
using dynamic payloads.

\begin{table*}[htb]
\centering
{\footnotesize
\begin{tabular}{|l|p{2.8cm}|l|l|} \hline \hline
{\bf MD5} & {\bf Dynamic Payload} & {\bf Description} & {\bf Malware Family} \\ \hline
34cb03276e426f8d61e782b8435d3147 & /assets/runme.png & ELF file to expoit root& GinMaster \\
a24d2ae57c3cee1cf3298c856a917100 & /assets/gbfm.png\newline/assets/install.png\newline/assets/installsoft.png\newline/assets/runme.png & ELF file to expoit root& GinMaster  \\
9e847c9a27dc9898825f466ea00dac81 & /assets/gbfa.png \newline /assets/install.png& ELF file to expoit root& GinMaster \\
 dcbe11e5f3b82ce891b793ea40e4975e& plankton\_v0.0.4.jar & download in runtime & Plankton\\
\hline \hline
\end{tabular}
\caption{Examples of Dynamic Payloads}
\label{table:dynamic_payload}
}
\end{table*}

\section{{\bf Analytic Capability of DroidAnalytics}} \label{sec: analytic_capability}

We conduct three experiments
and show how analysts can study malware, carry out
similarity measurement between applications, as well as
perform class association
among 150,368 mobile applications in the database.

\noindent
{\bf A. Detailed Analysis on Malware:}
Using DroidAnalytics,
analysts
can also discover which class or method uses suspicious API calls
via the {\em permission recursion} technique.

\begin{figure*}[htb]
\begin{center}
\fcolorbox{gray}{white}{
\includegraphics[width=0.8\textwidth]{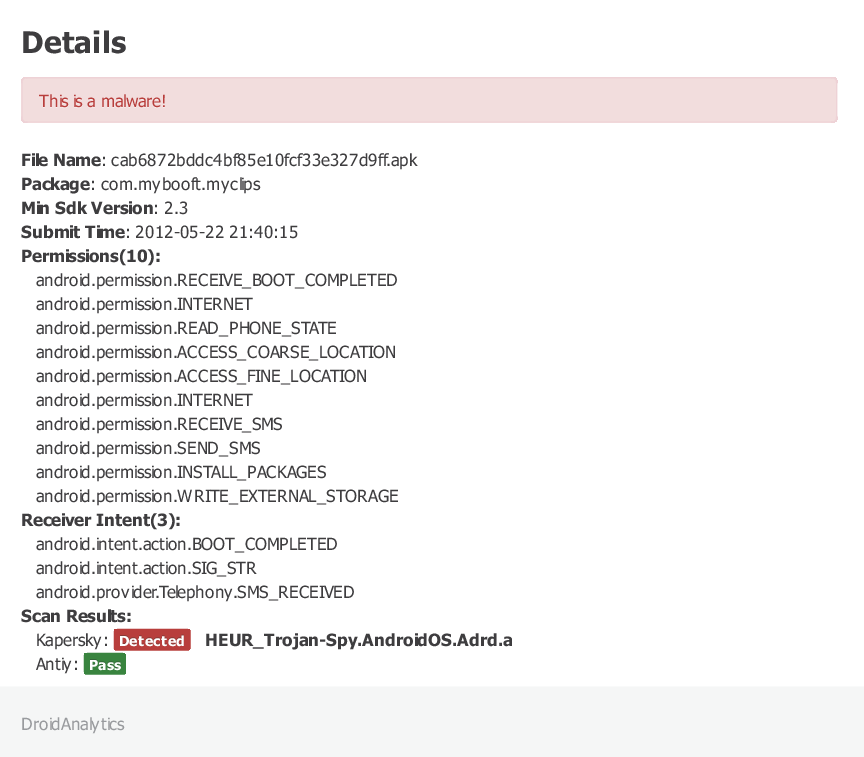}
}
\end{center}
\caption{Screen Capture of Common Analytics on Malware}
\label{Fig:app_detail}
\end{figure*}

\begin{figure*}[htb]
\begin{center}
\fcolorbox{gray}{white}{
\includegraphics[width=0.8\textwidth]{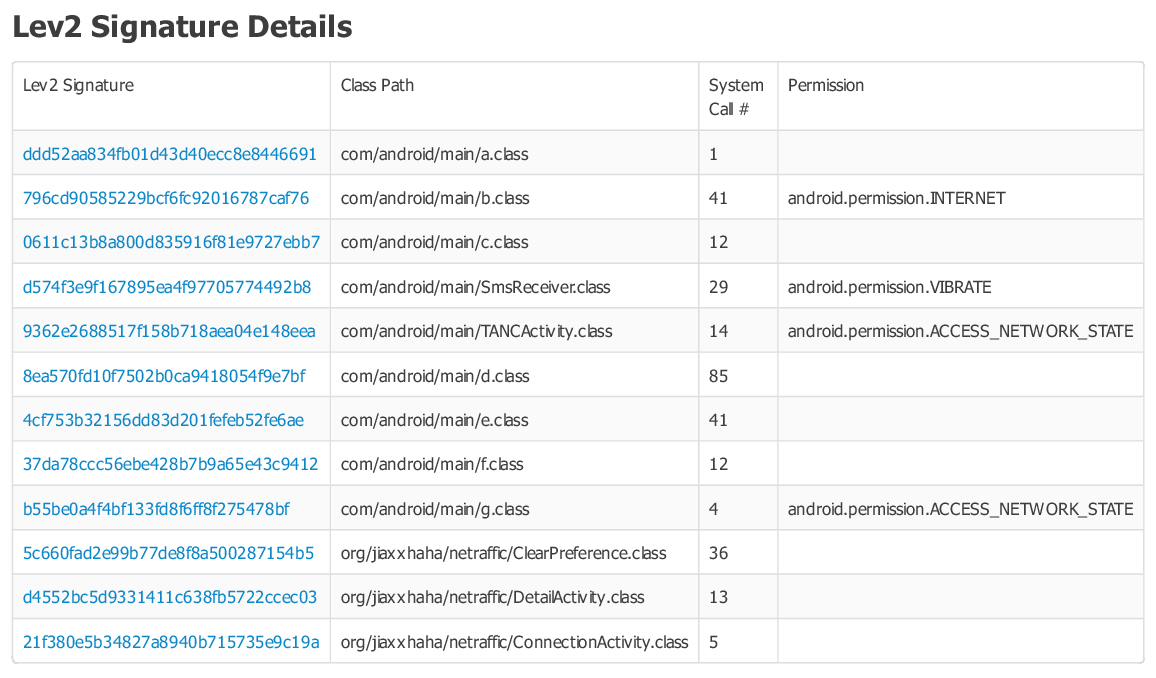}
}
\end{center}
\caption{Screen Capture of Detailed Lev2 Signature}
\label{Fig:lev2_signature}
\end{figure*}

\noindent
{\bf $\bullet$ Common Analytics on Malware.}
First, using the AIS parser, DroidAnalytics can reveal basic information of
an application like the cryptographic hash (i.e., MD5 value),
package name, broadcast receiver, $\ldots$, etc.
This is illustrated in Figure \ref{Fig:app_detail}.
In addition, DroidAnalytics has a built-in cloud-based APK scanner
that supports diverse anti-virus scan results (e.g., Kaspersky and Antiy)
to help analysts for reference.
Our cloud-based APK scanner is {\em extensible} to accommodate other
anti-virus scan engines.
Last but not least, DroidAnalytics can disassemble
the {\tt .dex} file and extract class number, method number,
and API calls number in each application, class or method.
These functionalities are useful so analysts can quickly
zoom in to the meta-information of a suspicious malware.
Figure \ref{Fig:lev2_signature} shows these functionalities.

\noindent
{\bf $\bullet$ Permission Recursion.}
Current state-of-the-art systems examine
the {\tt AndroidManifest.xml} to discover permissions of an application.
This is not informative enough since
analysts do not know which class or which method uses these permissions
for suspicious activities.
In DroidAnalytics, we can
discover the permission within a class or a method.
Since each permission is related to some API calls~\cite{felt:apd}.
In DroidAnalytics, we
tag permission to API calls in each method.
We combine the method permissions within the same class as class permission,
and combine all class permissions as application permission.
This helps analysts to quickly discover suspicious methods or classes.

\noindent
{\bf $-$ Experiment.}
In this experiment, we choose a popular malware family {\sf Kungfu}
and examine the permissions at the application/class/method levels.
Malware in the {\sf Kungfu} family can obtain
user's information such as IMEI number, phone model,..., etc.
It can also exploit the device and gain root privilege access.
Once the malware obtained the root level access, it installs
malicious application in the background as a back-door service.

\begin{figure*}
\centering
\includegraphics[width=360pt]{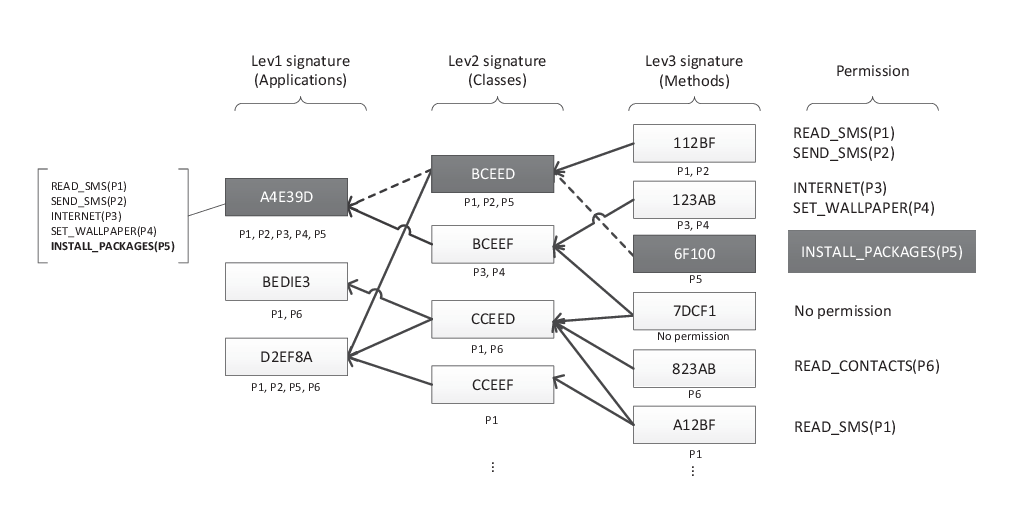}
\caption{Illustration of Detail Analytics on Malware}
\label{Fig:information}
\end{figure*}

We use DroidAnalytics to generate all three-level signatures.
Figure~\ref{Fig:information} shows the partial structure
of the signatures with permission
recursion of two  {\sf Kungfu} malware: {\tt A4E39D} and {\tt D2EF8A},
and together with a legitimate application {\tt BEDIE3} (Lev1 signature).
Firstly, based on the malware reports by our cloud APK scanner, we identify
{\tt A4E39D} and {\tt D2EF8A} are malware which come
from the {\sf Kungfu} family with different package names,
{\tt net.atools.android.cmwrap} and {\tt com.atools.netTrafficStats}
respectively.
By analyzing the Lev2 signature, we discover that {\tt BCEED} is
the common class which is within the two {\sf Kungfu} applications.
Secondly, from the Lev3 signature of {\tt BCEED},
we use  the permission recursion method to
expose the method {\tt 6F100} with the {\tt INSTALL\_PACKAGES} permission.
Note that the {\tt INSTALL\_PACKAGES} permission is a system permission
which the application can install other unrelated applications.
This shows how DroidAnalytics helps analysts to quickly
discover the malicious code of methods and classes of
the {\sf Kungfu} family.


\noindent
{\bf $\bullet$ Similarity Measurement.}
DroidAnalytics can also calculate the similarity between
two Android mobile applications, and the computation
is based on the three level signatures that
we discussed.
By comparing the similarity of applications,
we can determine whether an application is repackaged malware.
Moreover, because we use Lev2 (class) signature as the basic
building block, our approach not only can provide the similarity scores
between two applications, but can inform analysts the common and
different code segments between two applications.
This great facilitates analysts to carry out detailed analysis.

Our similarity score is based on the Lev2 signatures
which represent all classes within an application.
The similarity score is based on the Jaccard similarity coefficient.
Given $f_{a}$ and $f_{b}$ as two Lev2 signature sets
of two applications $a$ and $b$ respectively,
$f_a \cap f_b$
refers to the same Lev2 signatures of $a$ and $b$
(or the {\em common classes} of these two applications),
while $f_a \cup f_b$ represents
the set of classes of these two applications.
$S(x)$ is a function which returns
the total number of API calls in the set $x$.
Our similarity score equation between two applications is:
\begin{equation}
J_{app}(f_{a}, f_{b})=\frac{S( f_{a}\cap f_{b})}{S(f_{a}\cup f_{b})}.
\end{equation}

Let us how to use this similarity score to carry out analysis.
First, given a legitimate application $X$, we can find all
applications $\{Y_{1}, Y_{2},..., Y_{k}\}$ which are the top
$p\%$ (say $p=20$) applications that are similar to $X$.
Second, in the procedure of calculating
similarity, we can identify the common and different Lev2 signatures.
As a result, repackaged or mutated malware (i.e., $Y_i$),
can be easily detected
using the similarity score and the classes of
malicious behavior can be easily determined.

\noindent
{\bf $-$ Experiment.} We carry out the experiment using
similarity measurement based on the Lev2 and Lev3 signatures.
We use a benign application and calculate the similarity with
other applications in our database to find all repackaged malware of
this benign application. Furthermore, we can discover the differences
between this benign application and repackaged malware (at the code
level) to see what malicious codes are repackaged into legitimate application.

We choose a legitimate application called {\em ``Touch alarm''} which
is downloaded from Google official market. We use
DroidAnalytics to compute the similar scores with
malware in the database, Table \ref{table:scores} shows
the similarity scores between {\em ``Touch alarm''} and other
applications in our database (in decreasing order of similarity).  From the
table, we can clearly observe that the package names
of the top six applications are
the same: {\tt com.k99k.keel.touch.alert.freeze}.
The first one in the table is the legitimate {\em ``Touch alarm'' }application,
while the following five applications are repackaged malware
of {\sf Adrd}.
{\sf Adrd} steals the personal information like IMEI, hardware information
of users, and
it will encrypt the stolen information and upload to some remote server.
Moreover, it may dynamically download the
newest version of {\sf Adrd} and update itself.
The table shows that the malware writer inserted malicious codes
in benign {\em ``Touch alarm''}
and repackaged it to different variants of {\sf Adrd}.
\begin{table*}
\centering
{\scriptsize
\begin{tabular}{|l|c|c|r|r|} \hline \hline
{\bf MD5} & {\bf Package Name} & $\boldsymbol{S(f_{a}\cap f_{b})/S(f_{a}\cup f_{b})}$ & {\bf Similarity Score} & {\bf Detection Result} \\ \hline
{\tt 278859faa5906bedb81d9e204283153f} & com.k99k.keel.touch.alert.freeze & N/A & 1 & Not a Malware \\
{\tt effb70ccb47e8148b010675ad870c053} & com.k99k.keel.touch.alert.freeze & 674/878 & 0.76 & Adrd.w \\
{\tt ef46ed2998ee540f96aaa1676993acca} & com.k99k.keel.touch.alert.freeze & 674/878 & 0.76 & Adrd.w \\
{\tt cd6f6beff21d4fe5caa69fb9ff54b2c1} & com.k99k.keel.touch.alert.freeze & 674/878 & 0.76 & Adrd.w \\
{\tt 99f4111a1746940476e6eb4350d242f2} & com.k99k.keel.touch.alert.freeze & 674/884 & 0.77 & Adrd.a \\
{\tt 49bbfa29c9a109fff7fef1aa5405b47b} & com.k99k.keel.touch.alert.freeze & 674/884 & 0.77 & Adrd.a \\
{\tt 39ef06ad651c2acc290c05e4d1129d9b} & org.nwhy.WhackAMole & 332/674 & 0.49 & Adrd.cw \\
\vdots & \vdots & \vdots & \vdots & \vdots \\
\hline \hline
\end{tabular}
\caption{Similarity Scores of ``{\tt Touch Alarm}'' and other applications}
\label{table:scores}
}
\end{table*}

By using similarity measurement based on the three-level signature, DroidAnalytics
reveals the difference between two applications at the code level.
Table~\ref{table:CompareSin} shows the comparison of
legitimate application {\em ``Touch alarm''}
and repackaged malware.
Highlighted rows are the different level 2 signatures of the two
applications, while the other rows
represent same signature of common classes.
This shows that DroidAnalytics can easily identify different classes
of the two applications. By using permission recursion which
we discussed previously,
analysts can discover that the different level 2
signature {\tt e48040acb2d761fedfa0e9786dd2f3c2}
has READ\_PHONE\_STATE, READ\_CONTACTS and SEND\_SMS in repackaged malware.
Using DroidAnalytics for further analysis,
we find suspicious API calls like
{\tt android/telephony/TelephonyManager;->\\getSimSerialNumber},
{\tt android/telephony/gsm/\\SmsManager;->endTextMessage} and
{\tt android/telephony/TelephonyManager;->\\getDeviceID}.
Last but not least, we
determine the malware writer inserted these malicious codes
into legitimate application and published repackaged malware
to various third party markets.

\begin{table*}
\centering
{\tiny
\begin{tabular}{l@{}cp{1.3cm}rl@{}cp{2.5cm}} \hline \hline
\multicolumn{3}{c}{{\bf Application A} (MD5: {\tt 278859faa5906bedb81d9e204283153f})} &\phantom{a}& \multicolumn{3}{c}{{\bf Repackaged Malware of A} (MD5: {\tt effb70ccb47e8148b010675ad870c053})} \\
\cmidrule{1-3} \cmidrule{5-7}
Level 2 Signature & \# of API calls  & Permission & & Level 2 Signature & \# of API calls& Permission \\ \hline
{\tt 02bcaaa836d530035bb8db801c85cffd} & 17 & N/A & & {\tt 02bcaaa836d530035bb8db801c85cffd} & 17 & N/A \\
{\tt 0611c13b8a800d835916f81e9727ebb7} & 12 & N/A & & {\tt 0611c13b8a800d835916f81e9727ebb7} & 12 & N/A \\
\rowcolor{light-gray}&  &  & & {\tt 178e2481067b194e00187cfaadb12f4d} & 1 & N/A \\
{\tt 6720117047d39c2e2e90fa7e896e1615} & 22 & WAKE\_LOCK & & {\tt 6720117047d39c2e2e90fa7e896e1615} & 22 & WAKE\_LOCK \\
\rowcolor{light-gray}&  &  & & {\tt e48040acb2d761fedfa0e9786dd2f3c2} & 62 & READ\_PHONE\_STATE\newline READ\_CONTACTS\newline SEND\_SMS  \\
\vdots & \vdots & \vdots & & \vdots & \vdots & \vdots \\
\hline \hline
\end{tabular}
\caption{Comparison Between Legitimate App
``{\tt Touch Alarm}'' and Repackaged Malware}
\label{table:CompareSin}
}
\end{table*}

\noindent
{\bf $\bullet$ Class Association.}
Traditional analysis on malware only focuses on one malware but
can not associate malware with other malware or applications.
DroidAnalytics can associate
legitimate applications and other malware in the class level and/or
method level. Given a class signature (or Lev2 signature), DroidAnalytics
keeps track of how many legitimate applications or malware using this particular class.
Also, with the methodology of permission recursion,
DroidAnalytics can indicate the permission usage of this
class signature. By using class association, we can easily
determine which class or method may possess malicious behavior,
and which class is used for common task, say for pushing advertisement.
Lastly, for class signatures which are used by many known malware but zero
legitimate application, these are classes that analysts need to pay
special attention to because it is very likely that
they contain malicious code and are used in many repackaged or
obfuscated malware.

\begin{figure*}[htb]
\begin{center}
\fcolorbox{gray}{white}{
\includegraphics[width=0.8\textwidth]{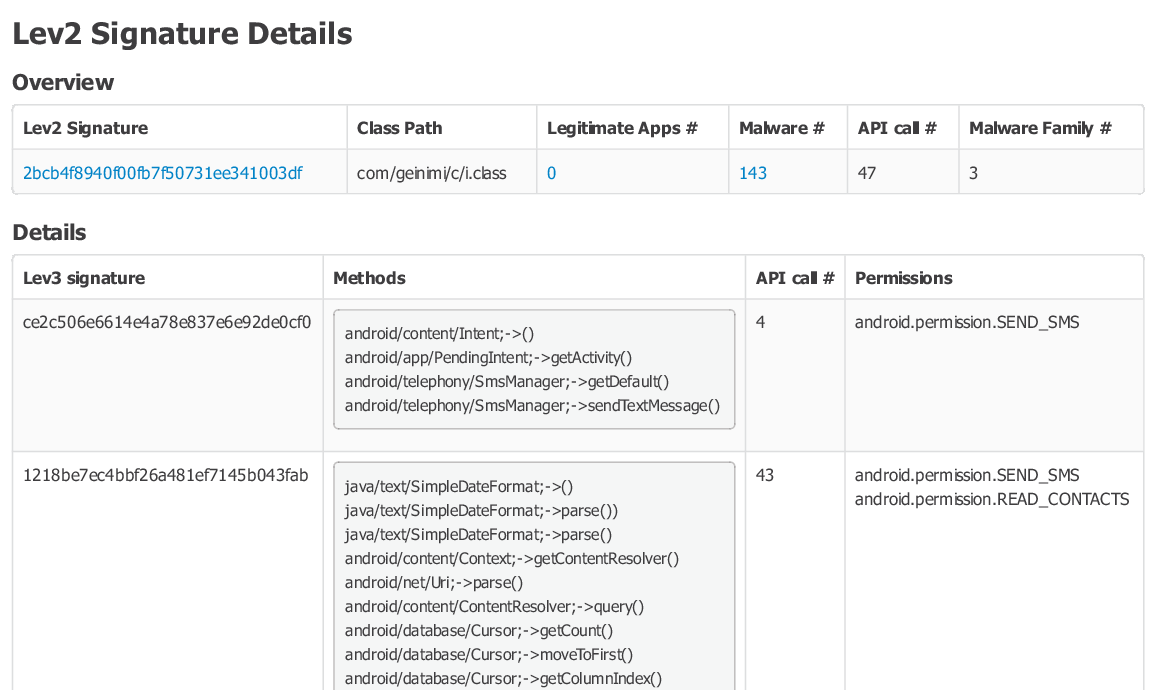}
}
\end{center}
\caption{Detailed Lev2 Signature of Repackaged Geinimi Malware}
\label{Fig:lev2_details}
\end{figure*}

\begin{figure*}[htb]
\begin{center}
\fcolorbox{gray}{white}{
\includegraphics[width=0.8\textwidth]{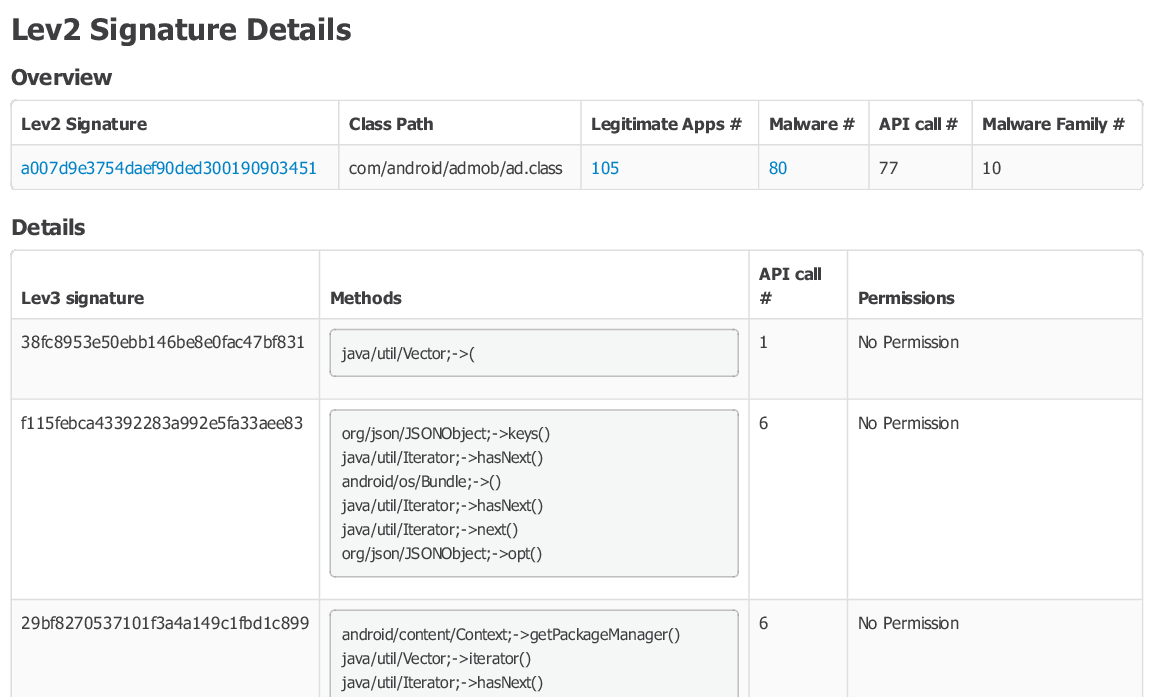}
}
\end{center}
\caption{Detailed Lev2 Signature of Advertisement Library}
\label{Fig:lev2_details_ad}
\end{figure*}

\noindent
{\bf $-$ Experiment.}
Using DroidAnalytics, we carry out the class association experiment using
1,000 legitimate appliactions and 1,000
malware as reported by Kaspersky.
Figure \ref{Fig:lev2_details} illustrates the results.
After the class association, we discover
a class (Lev2) signature {\tt 2bcb4f8940f00fb7f50731ee341003df}
which is used by 143 malware and zero legitimate application.
In addition, the 143 malware are all from the {\sf Geinimi} family. Furthermore,
this class has 47 API calls and uses
{\tt READ\_CONTACTS} and {\tt SEND\_SMS} permissions.
Therefore, we quickly identify this class contains malicious codes.
In the signature database, we also find a class (Lev2) signature
{\tt 9067f7292650ba0b5c725165111ef04e} which is used by 80 legitimate
applications and 42 malware.
Further analysis shows that this class is used by
similar number of legitimate applications and malware, and
this class uses an advertisement library called DOMOB\cite{domob}.
Another class (Lev2) signature {\tt a007d9e3754daef90ded300190903451}
is used by 105 legitimate applications
and 80 malware. Further examination shows that it is a class from
the Google official library called AdMob\cite{admob}.
This is illustrated in Figure \ref{Fig:lev2_details_ad}.

In summary, all experimented samples are presented
in Table \ref{table:results1}, which represents
the known malware in our database.
The detection results is based on the cloud anti-virus engine using
various detection engines (i.e.,
Kaspersky and Antiy(linux version)).
Note that in the last column, $R$ ($G$) represents
repackaged (generic) malware family.
For those malware families with less than five samples,
we lump them as ``others'' in the table.
Antiy (linux version) \cite{antisc} is a commercial anti-virus product that we
obtained from the company, and the product is known to run
the same engine for detecting malware in smartphones.
The rows are sorted in alphabetical order. Highlighted rows show
the common malware families detected by both Kaspersky and Antiy.
Others are uniquely detected by one anti-virus product.
The penultimate row shows there are 1,295 common malware samples detected by
these two anti-virus products.
Hence, the number of unique malware samples is 2,148.

\definecolor{light-gray}{gray}{0.93}
\begin{table}
\begin{center}
{\footnotesize
    \begin{tabular}{|llll|}
        \hline \hline
{\bf Kaspersky} & {\bf Samples} & {\bf Antiy} & {\bf Samples}
\\ \hline
\rowcolor{light-gray} Adrd & 176 & Adrd & 57\\
 &  & AnSer & 5\\
 &  & App2card & 8\\
\rowcolor{light-gray} BaseBrid & 611 & Keji(BaseBrid) & 299\\

\rowcolor{light-gray} CrWind & 5 & Crusewin(CrWind) & 5\\
 &  & Deduction & 19\\
DorDrae & 10 &  & \\

 &  & emagsoftware & 15\\
\rowcolor{light-gray} FakePlayer & 6 & FakePlayer & 6\\
Fatakr & 16 &  & \\
\rowcolor{light-gray} Fjcon & 142 & fjRece(Fjcon) & 141\\
\rowcolor{light-gray} Gapev & 5 & gapp(Gapev) & 4\\
\rowcolor{light-gray} Geinimi & 139 & Geinimi & 128\\
 &  & GingerBreak & 11\\
\rowcolor{light-gray} GinMaster & 31 & GingerMaster(GinMaster) & 26\\
\rowcolor{light-gray} Glodream & 13 & GoldDream & 7\\

Gonca & 5 &  & \\

 &  & i22hk & 5\\
 &  & Itfunz & 10\\

 &  & jxtheme & 13\\
\rowcolor{light-gray} Kmin & 192 & Kmin & 179\\
\rowcolor{light-gray} KungFu & 144 & KungFu & 78\\
 &  & Lightdd & 7\\
\rowcolor{light-gray} Lotoor & 113 & Lotoor & 15\\

 &  & MainService & 121\\
\rowcolor{light-gray} MobileTx & 15 & tianxia(MobileTx) & 14\\
Nyleaker & 7 &  & \\
Opfake & 8 &  & \\
\rowcolor{light-gray} Plangton & 126 & Plankton(Plangton) & 1\\
\rowcolor{light-gray} Rooter & 26 & DroidDream(Rooter) & 17\\
 &  & RootSystemTools & 7\\
\rowcolor{light-gray} SeaWeth & 11 & seaweed(SeaWeth) & 11\\
\rowcolor{light-gray} SendPay & 9  & go108(SendPay) & 10\\
\rowcolor{light-gray} SerBG & 23 & Bgserv(SerBG) & 31\\

Stiniter & 50 &  & \\

 &  & Universalandroot & 27\\
 &  & Latency & 28\\
 &  & Visionaryplus & 1\\
 &  & Wukong & 10\\
\rowcolor{light-gray} Xsider & 14 & jSMSHider(Xsider) & 8\\
 &  & YouBobmg & 20\\
Yzhc & 35 &  & \\
 &  & Z4root & 47\\

 &  & Zft & 5\\
 Others (42 families) & 71 & Others (27 families) & 44 \\
\rowcolor{light-gray} Common & 1295 & Common & 1295 \\
All & 2003 & All & 1440 \\
\hline \hline
    \end{tabular}
	\caption{Detection Results of our Cloud-based Apk Scanner}
	\label{table:results1}
}
\end{center}
\end{table}

\section{{\bf Zero-day Malware Detection}} \label{sec: zeroday}

Here, we show a novel methodology in using DroidAnalytics
to detect the zero-day repackaged malware.
We analyze three zero-day
malware families to illustrate the effectiveness of our system.

\subsection{Zero-day Malware}
Zero-day malware is a new malware that
current commercial
anti-virus systems cannot detect.
Anti-virus software usually relies on signatures to identify malware.
However, signature can only be generated when samples are obtained.
It is always a challenge for anti-virus companies to
detect the zero-day malware, then
update their malware detection engines as quickly as possible.

In this paper, we define an application as a zero-day malware
if it has malicious behavior and it cannot be detected by
popular anti-virus software (e.g.,
Kaspersky, NOD32, Norton) using their latest signature database.
As of November, 2012, we use
DroidAnalytics and have successfully detected 342
zero-day repackaged malware in five different families:
{\sf AisRs}, {\sf Aseiei}, {\sf AIProvider}, {\sf G3app},
{\sf GSmstracker} and {\sf YFontmaster}
(please refer to Table \ref{table:zero-day} for reference).
In this paper, we use the name of the injected package
(not the name of the repackaged applications)
as the name of its malware family. Furthermore, all samples are scanned by
Kaspersky, NOD32, Norton and Antiy using their latest
database in November, 2012. We also uploaded these samples to the
virustotal\cite{virustotal} for malware detection analysis.
Note that none of the submitted samples was reported as
a malware by these engines when we carried out our experiments.

\begin{table}[htb]
\centering
{\footnotesize
\begin{tabular}{|l|l|l|} \hline \hline
{\bf Family Name} & {\bf Number} & {\bf Malicious Package Name} \\ \hline
AisRs      & 87             & com.ais.rs\\
Aseiei     & 64             & com.aseiei\\
AIProvider & 51     & com.android.internal.provider\\
G3app      & 96     & com.g3app\\
GSmstracker     & 10     & com.gizmoquip.smstracker\\
YFontmaster     &  34    & com.yy.fontmaster \\ \hline
{\bf All}     & 342     & \\
\hline \hline
\end{tabular}
\caption{Zero-day Repackaged Malware Samples}
\label{table:zero-day}
}
\end{table}

In \cite{Abhi:repackaged,Gary:repackaged}, authors reported that
nearly 86.0\% of all Android malware
are actually repackaged versions of some legitimate applications.
By camouflaging to some legitimate applications,
repackaged malware can easily deceive users.
Given the large percentage of repackaged malware, we explore
the effectiveness of using DroidAnalytics
to detect the {\em zero-day repackaged malware}.

\subsection{Zero-Day Malware Detection Methodology}
The process of detecting zero-day repackaged malware can be summarized
by the following steps.

\noindent
{\bf Step 1:} We first construct a white list for common and legitimate classes.
For example, we add all legitimate level 2 signatures,
such as those in utility libraries
(e.g., Json library) or advertisement libraries
(e.g., Google Admob library, Airpush library)
to the white list. All level 2 signatures in the
white list will not be used to calculate the
similarity score between two applications.

\noindent
{\bf Step 2:}
We calculate the number of common API calls between two given
applications in the database. This can be achieved
by using the similarity score
in Equation (\ref{equation: similarity score}).

\begin{equation}
S_{app}(f_{a}, f_{b})=S( f_{a}\cap f_{b}),
\label{equation: similarity score}
\end{equation}
where $f_{a}$ and $f_{b}$ as two level 2 signature sets
of two applications $a$ and $b$ respectively,
$S(x)$ is a function to indicate
the total number of API calls in the set $x$.
The above similarity score between two
repackaged malwares focus on
the {\em common} repackaged API calls
that correspond to the malicious logic,
and ignore the effect of other API calls in these two applications.

\noindent
{\bf Step 3:} Assume we have $N$ applications in the database,
then we start with $N$ clusters.  The distance between two
clusters is the similarity score we mentioned in Step 2.
We select two applications which have the largest similarity
score and combine them into one cluster. For this new
cluster, we re-calculate the similarity score between this new
cluster with other $N-2$ clusters. The new similarity
score is computed by averaging
all similarity scores between all applications in two different clusters.

\noindent
{\bf Step 4:} Again, we combine two clusters which have the
largest similarity score.  We continue this step until the similarity
score between any two clusters is less than a pre-defined threshold $T$
(say $T$ is 100).

\noindent
{\bf Step 5:}
After we finish the clustering process, we will use anti-virus engines
to scan all of these $N$ applications.
Each application may be classified as {\em legitimate} or {\em malicious}.

\noindent
{\bf Step 6:}
If a cluster has more than $n$ applications  (say $n=10$)
and a small fraction $f$ (say $f \leq 0.2$) is classified as malicious.  This
should be a {\em suspicious cluster} since
it is very unlikely that more than $n$ applications
are similar (in terms of class functionality) in the real-world,
and most of them are classified as benign.
Hence, the similarity comes when some of these applications
are repackaged.  We then extract their common classes
(using our level 2 signature) and examine these classes.
Once we find any malicious logic in these common classes,
we discover a zero-day repackaged malware family.

\noindent
{\bf Experimental results in discovering zero-day repackaged malware:}
Let us present the results of using DroidAnalytics to discover three
zero-day repackaged malware families.

\begin{figure}[htb]
\begin{center}
\includegraphics[width=0.45\textwidth]{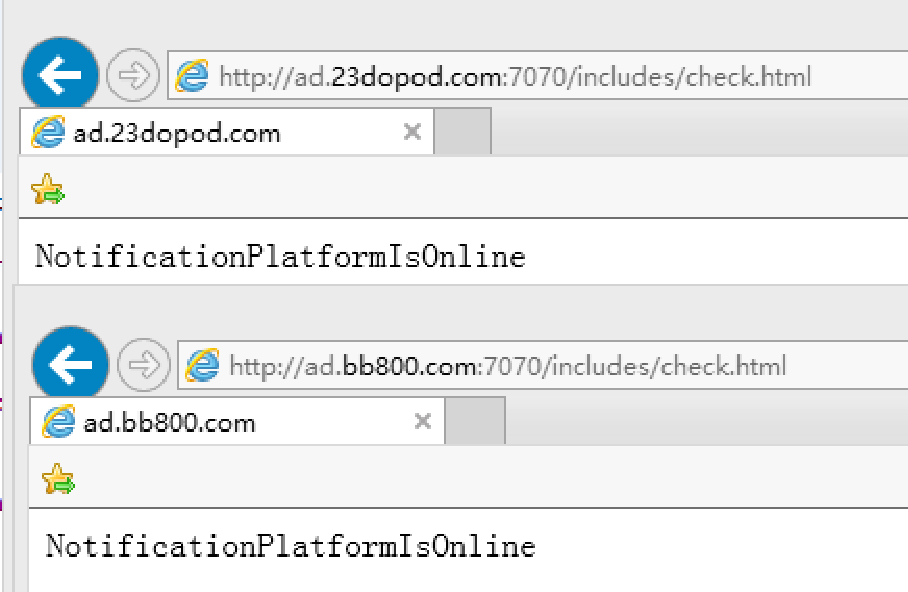}
\end{center}
\caption{Remote Servers of AisRs}
\label{Fig:AisRs_1}
\end{figure}

\begin{figure}[htb]
\begin{center}
\includegraphics[width=0.45\textwidth]{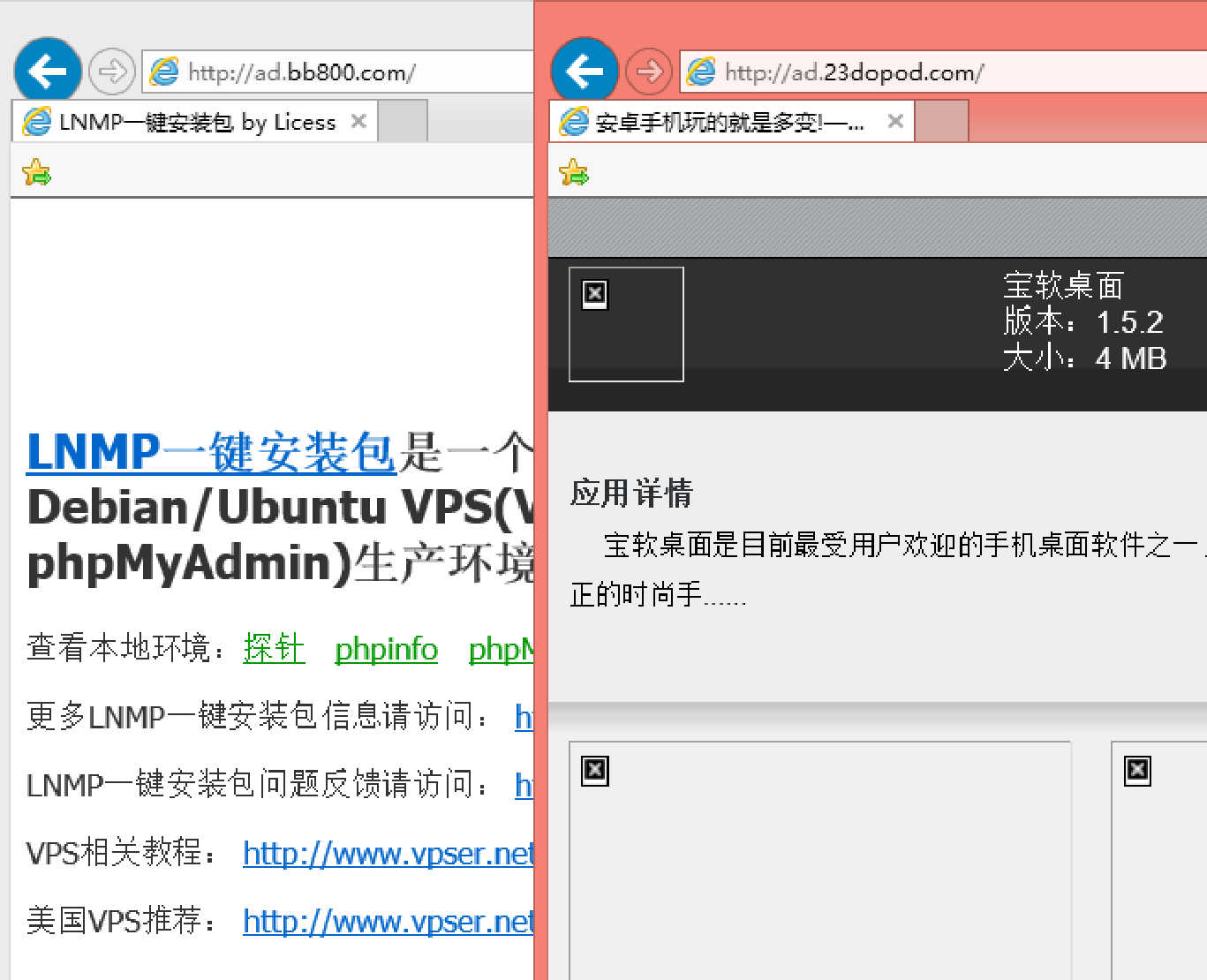}
\end{center}
\caption{Camouflaged Software Download Websites}
\label{Fig:AisRs_2}
\end{figure}

\noindent
{\bf - AisRs family:} We discover 87 samples of {\sf AisRs}
family in our database. All the malware are repackaged from
legitimate applications (e.g., {\em com.fengle.jumptiger},
{\em com.mine.videoplayer})
and all of them have
a common malicious package named ``{\tt com.ais.rs}''.
This malware contains a number of botnet commands that
can be remotely invoked. When the malware runs, it will first
communicate with two remote servers (see Figure \ref{Fig:AisRs_1}).
These two servers are camouflaged as software download
websites(see Figure \ref{Fig:AisRs_2}).
If any of these two servers is
online, the malware will receive some commands like
downloading other {\tt .apk} files. They are not necessary malware,
but they contain advertisement from
the website, ``http://push.aandroid.net''.
It is interesting to note that the address
is ``{\bf a}android.net'', not ``android.net''.
Also, one of the many botnet commands is to save the user's application
installation lists and system information to the {\tt .SQLite}
file, then upload this file to a remote server.

\begin{figure}[htb]
\begin{center}
\includegraphics[width=0.45\textwidth]{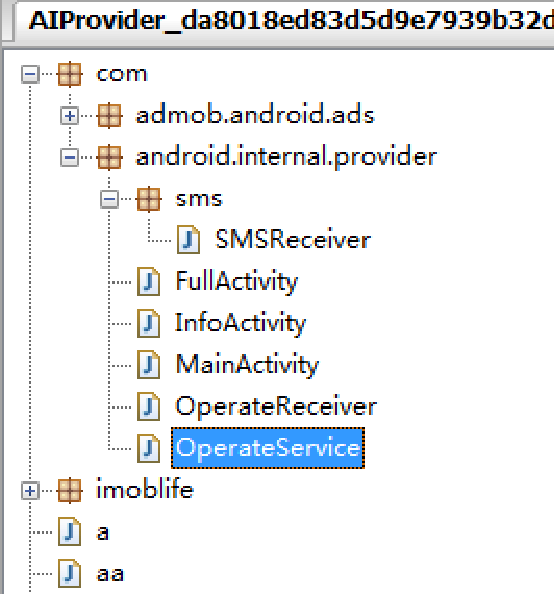}
\end{center}
\caption{Disassembled Repackaged Codes of AIProvider}
\label{Fig:AIProvider_1}
\end{figure}

\begin{figure}[htb]
\begin{center}
\includegraphics[width=0.30\textwidth]{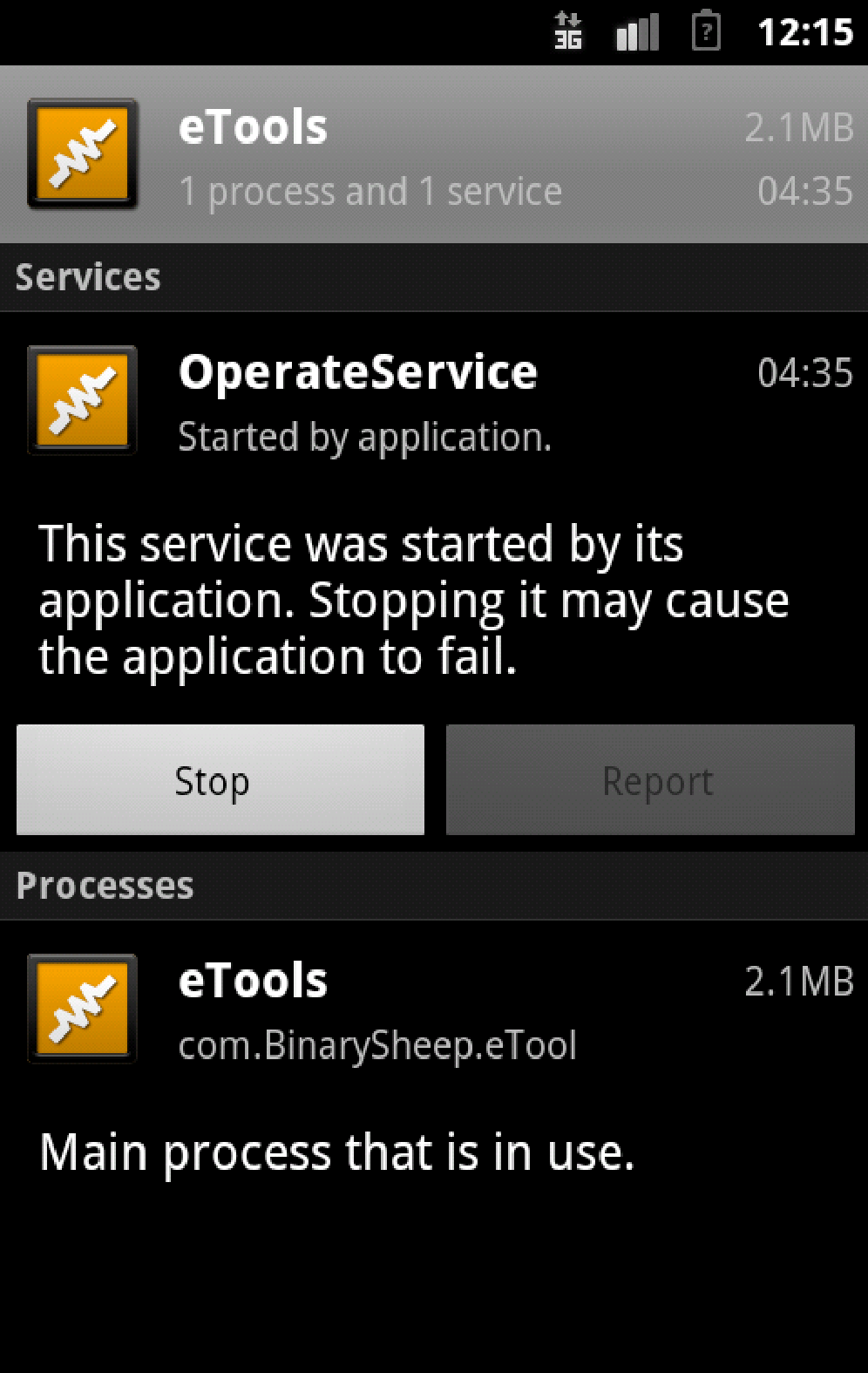}
\end{center}
\caption{Malicious OperateService of AIProvider}
\label{Fig:AIProvider_2}
\end{figure}

\noindent
{\bf - AIProvider family:} We discover 51 samples of {\sf AIProvider}
family in our database. All the malware are repackaged from
legitimate applications (e.g., {\em jinho.eye\_check},
{\em com.otiasj.androradio}) and
all of them have a common malicious package
named ``{\tt com.android.internal.provider}''. There are several
interesting characteristics of this malware. Firstly, the
malicious package name is disguised as a system package name.
Since DroidAnalytics does not detect malware based on the package name,
so our system can easily discover this repackaged malware. Secondly,
this malware uses DES to encrypt all SMS
information (e.g., telephone numbers, SMS content) and store them
in the {\tt DESUtils} class. Thirdly, the malware will start a
service called {\tt OperateService} in the
background when it receives ``{\tt BOOT\_COMPLETED}''
broadcast(see Figure \ref{Fig:AIProvider_1} and \ref{Fig:AIProvider_2}).
This service will decrypt the SMS information in the
{\tt DESUtils} class and use this information to send SMS messages
without any notification.

\begin{figure}[htb]
\begin{center}
\includegraphics[width=0.30\textwidth]{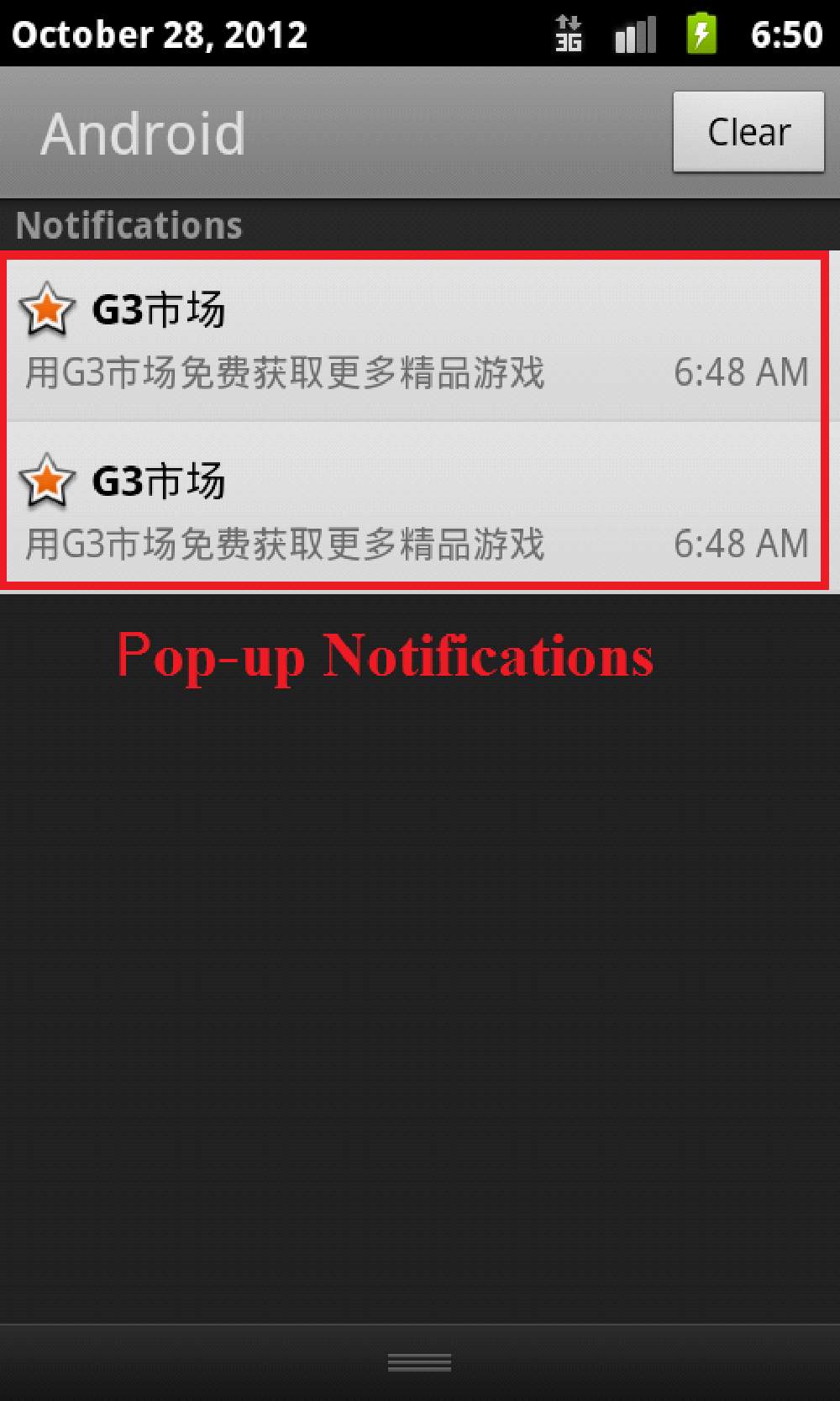}
\end{center}
\caption{Trigger Notification of G3app}
\label{Fig:g3app_2}
\end{figure}

\begin{figure}[htb]
\begin{center}
\includegraphics[width=0.30\textwidth]{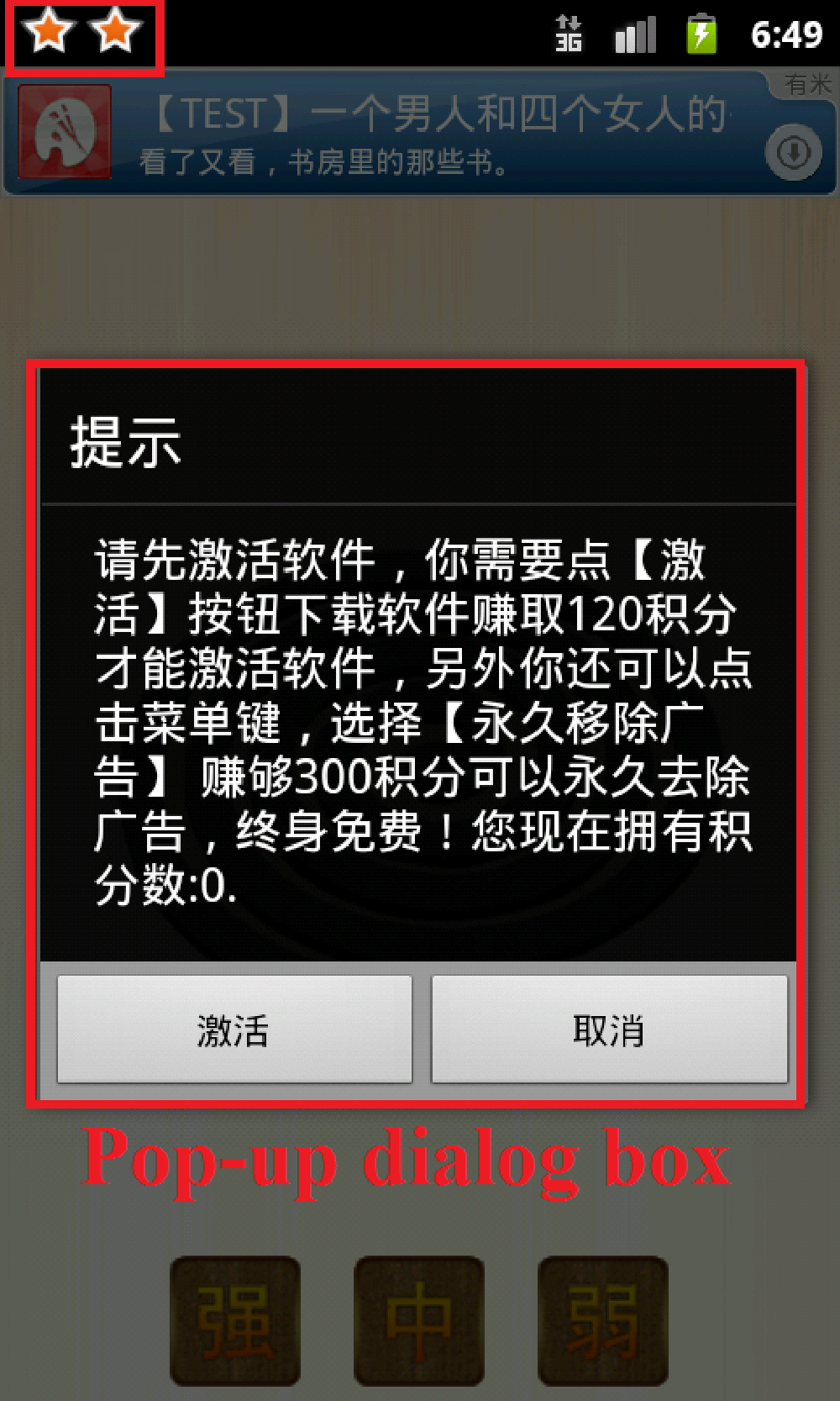}
\end{center}
\caption{Trigger Button of G3app}
\label{Fig:g3app_1}
\end{figure}

\noindent
{\bf - G3app family:} We discover 96 samples of {\sf G3app}
family in our database. All the malware are repackaged from
legitimate applications (e.g., {\em com.openg3.virtua lIncomingCall},
{\em com.cs.android.wc3}) and
all of them have a common malicious package
named ``{\tt com.g3app}''. There are several
malicious behaviors in the {\sf G3app} malware family. Firstly, the malware
will frequently pop up notification on the status bar and entice
users to select it (see Figure \ref{Fig:g3app_2}).
Secondly, the malware will inject trigger codes to
every button of the legitimate application(see Figure \ref{Fig:g3app_1}).
If the user presses any
button in the repackaged application or the notification in the status bar,
the malware will download other applications
from the remote server. We believe the
hackers want to use repackaged malware to publicize their applications and
use these advertisements for financial gain.


\section{{\bf Related Work}} \label{sec: related}

Before the rapid increase of Android malware in 2011,
researchers focused on the permission and capability leaks of
Android applications. E.g.,
David Barrera {\em et al.}\cite{barrera:mmap}
propose a methodology to explore and analyze
permission-based models in Android.
Stowaway\cite{felt:apd} is a tool for detecting permission
over privilege in Android,
while ComDroid\cite{felt:com} is a tool which detects communication
vulnerabilities. Woodpecker\cite{Michael:2012} analyzes each
application on a smartphone to explore the
reachability of a dangerous permission from a public,
unguarded interface. William Enck {\em et al.} \cite{Enck:2009}
propose a lightweight mobile phone application with a certification-based
permission.
In this paper, instead of malware detection,
we focus on designing an analytic system that helps analysts
to dissect, analyze and correlate Android malware
with other Android applications in the database.
We propose a novel signature system
to identify malicious code segments and associate
with other malware in the database.
Our signature system is robust against
obfuscation techniques that hackers may use.

In August 2010, Kaspersky reported the first SMS Trojan,
known as {\sf FakePlayer} in Android systems\cite{Carlos:2011}.
Since then, many malware and their variants have been discovered and
mobile malware rapidly became a serious threat.
Felt {\em et al.} study 18 Android malware in \cite{felt:smmw}.
Enck {\em et al.} \cite{enck:saas} carry out a study with
1,100 android applications but no malware was found.
Zhou Yang {\em et al.}\cite{jiang:oakland} study characterization and
evolution of Android malware with 1,260 samples. However, they did not show how to
systematically collect, analyze and correlate these samples.

Yajin Zhou {\em et al.}\cite{zhou:drsa} were the first to present
a systematic study for the detection of malicious applications on
Android Markets. They successfully discover 211 malware.
Their system, DroidRanger, needs malware samples to
extract the footprint of each malware family before the
known malware detection.
For zero day malware,
DroidRanger serves as a filtering system.
After the filtering process, suspicious malware needs to be
manually analyzed.
DroidMOSS\cite{jiang:moss} is an Android system to detect
repackaged applications using fuzzy hashing.
As stated in \cite{jiang:moss}, the system is not designed for
general malware detection.
Furthermore, the similarity score provided by
DoridMoss is not helpful in malware analysis.
In addition, obfuscation techniques can change the order of classes and
methods execution, and this will introduce large deviation
in the measure used in DroidMOSS.
Michael Grace et al. develop
RiskRanker\cite{risk} to analyze whether a particular application exhibits
dangerous behavior.
It uses class path as the malware family feature to detect more mutations.
However,
obfuscation can easily rearrange the opcode along an execution path. So
using class path for malware feature is not effective under
obfuscation attack.
Our system overcomes these problems by using a novel
signature algorithm to extract the malware feature at
the opcode level so it captures the semantic
meaning for signature generation.

For PC based malware, a lot of research work focus on
the signature based malware detection.
For example, authors in \cite{Andreas07} discussed the limitation of
using signature to detect malware.
In \cite{okane:ohm}, authors described the obfuscation techniques to
hide malware.
Authors in \cite{Christodorescu:2007} presented an automatic system to
mine specifications of
malicious behavior in malware families.
Paolo Milani Comparetti {\em et al.}\cite{Comparetti:2010}
proposed a solution to determine malicious functionalities
of malware. However, mobile malware has different
features as compared with PC based malware. It is difficult to transform
a PC based malware detection solution for mobile devices.
For example, \cite{dimva12} reported that many anti-virus products
have poor performance in detecting Android malware mutations,
although these products performed reasonably well for PC based malware.
Repackaging is another characteristic of android malware.
Authors in \cite{jiang:moss} showed there are many repackaged applications in
Android third party markets and significant number of these applications
is malware. Based on the above studies, it is clear that
a more sophisticated methodology to detect and analyze
Android malware is needed.

\section{ {\bf Conclusion}} \label{section:conclusion}

We present DroidAnalytics, an Android malware analytic system
which can automatically collect malware,
generate signatures for applications, identify malicious code segment (even
at the opcode level), and at the same time, associate the malware
under study with various malware and applications in the database.
Our signature methodology provides
significant advantages over traditional cryptographic hash like MD5-based
signature.
We show how to use DroidAnalytics to quickly retrieve, associate and reveal
malicious logics.
Using the permission recursion technique
and class association, we show how to retrieve the permissions of
methods, classes and application (rather than
basic package information), and associate all applications in the opcode level.
Using DroidAnalytics, one can easily discover
repackaged applications via the similarity score.
Last but not least, we have used DroidAnalytics to
detect 2,494 malware samples
from 102 families, with 342 zero-day
malware samples from six different families.
We have conducted extensive experiments to
demonstrate the analytic and malware detection capabilities of DroidAnalytics.


\bibliographystyle{IEEEtran}
\bibliography{paper}

\end{document}